\documentclass[superscriptaddress,twocolumn,english,floatfix,aps,prd,amsmath,amssymb,longbibliography]{revtex4-1}
\usepackage[T1]{fontenc}
\usepackage[latin9]{inputenc}
\usepackage{babel}
\usepackage{polski}
\usepackage{color}
\usepackage{times}
\usepackage{epsfig}
\usepackage{subfigure}
\usepackage{amsmath}
\usepackage{amssymb}
\usepackage{graphicx}
\usepackage[colorlinks=true]{hyperref}
\hypersetup{citecolor=blue,linkcolor=blue,urlcolor=blue}

\begin{document}

\title{Relativistic bands in the discrete spectrum of created particles in an oscillating cavity}

\author{Danilo T. Alves}
\email{danilo@ufpa.br}
\affiliation{Faculdade de F\'{i}sica, Universidade Federal do Par\'{a}, 66075-110, Bel\'{e}m, Par\'{a}, Brazil}

\author{Edney R. Granhen}
\email{granhen@unifesspa.edu.br}
\affiliation{Instituto de Ci\^{e}ncias Exatas, Faculdade de F\'{i}sica, Universidade Federal do Sul e Sudeste do Par\'{a}, 68505-080, Marab\'{a}, PA,  Brazil}

\author{Jo\~{a}o Paulo da S. Alves}
\email{joao.alves@ifpa.edu.br}
\affiliation{Instituto Federal do Par\'{a}, 66093-020, Bel\'{e}m, PA,  Brazil}

\author{Williams A. Lima}
\email{williamslima@on.br}
\affiliation{Observat\'{o}rio Nacional, 20921-400, Rio de Janeiro, RJ,  Brazil}

\date{\today}

\begin{abstract}
We investigate the dynamical Casimir effect for a one-dimensional resonant cavity, 
with one oscillating mirror.
Specifically, we study the discrete spectrum of created particles in a region of frequencies above the oscillation frequency $\omega_0$
of the mirror.
We focus our investigation on an oscillation time equal to $2L_0/c$, where $L_0$ is the initial and final length of the cavity, and
$c$ is the speed of light.
For this oscillation duration, a field mode, after perturbed by the moving mirror, never meets this mirror in motion again,
which allows us to exclude this effect of re-interaction on the particle creation process.
Then, we describe the evolution of the particle creation with frequencies above $\omega_0$
only as a function of the relativistic aspect of the mirror's velocity.
In other words, we analyze the formation of relativistic bands in a discrete spectrum of created particles.

\end{abstract}

\maketitle

\section{Introduction}

Moore, in his pioneering paper on the dynamical Casimir effect (DCE) \cite{Moore-1970}, pointed out that
photons could be created by the excitation of the quantum vacuum by a moving mirror
(the DCE was also investigated in other pioneering articles
\cite{DeWitt-PhysRep-1975,Fulling-Davies-PRSA-1976,Davies-Fulling-PRSA-1977,Candelas-PRSA-1977} and 
excellent reviews can be found in the literature \cite{Dodonov-JPCS-2009, Dodonov-PhysScr-2010, Dalvit-CasimirPhysics-2011,Dodonov-Phys-2020}).
On the other hand, he remarked that the photon creation predicted by him was negligible 
to be detected experimentally \cite{Moore-1970}. 
One of the problems of the particle creation via DCE is that, under laboratory conditions,
the maximum velocity that an oscillating mirror can achieve is very small in comparison to the
speed of light \cite{Dodonov-Klimov-PRA-1996}.
To circumvent this problem, Dodonov and Klimov investigated
the possibility of observation of the DCE considering 
a gradual accumulation of photons in a resonant cavity,
so that a significant and measurable effect could be obtained \cite{Dodonov-Klimov-PRA-1996}.
Other several proposals have been made, focusing on the observation of the particle creation from vacuum by experiments based on
the mechanical oscillation of mirrors
\cite{Kim-PRL-2006,Brownell-JPA-2008,Motazedifard-2018,Sanz-Quantum-2018,Qin-PRA-2019,Butera-PRA-2019}, 
but the observation remains as a challenge \cite{Dodonov-Phys-2020}.

In a more general point of view, the particle creation from vacuum occurs when a quantized field is submitted 
to a time-dependent boundary condition. Therefore, a moving mirror exciting the vacuum is just a particular case. 
Yablonovitch \cite{Yablonovitch-PRL-1989} and Lozovik \textit{et al.} \cite{Lozovik-PZhETF-1995} 
proposed alternative ways to excite the quantum vacuum, by means of time-dependent boundary conditions imposed on a material medium.
Moreover, a motionless mirror whose internal properties rapidly vary in time can simulate a moving mirror.
Several experimental proposals emerged in this context
\cite{Braggio-EPL-2005, Braggio-JPA-2008, Braggio-JPCS-2009,Johansson-PRL-2009,Dezael-2010,Wilson-Nature-2011,Lahteenmaki-PNAS-2013,
Motazedifard-2015}. 
One of them led Wilson \textit{et al.} to observe experimentally the particle creation
from vacuum \cite{Wilson-Nature-2011}, getting a maximum effective velocity $v\approx 0.1c$.
Other experiments have also been done \cite{Lahteenmaki-PNAS-2013, Vezzoli-CommPhys-2019,Schneider-et-al-PRL-2020}, 
with one of them getting a maximum effective velocity $v \approx 0.31c$ \cite{Schneider-et-al-PRL-2020}.

In the present paper, we investigate aspects of the
problem combining a resonant oscillating cavity with a relativistic maximum velocity of its moving mirror.
Specifically, we study the discrete spectrum of created particles
in the context of a real massless scalar field in $(1+1)\text{D}$, inside a resonant cavity with one relativistic moving mirror, 
oscillating with a frequency $\omega_0$.
Moreover, we impose the Dirichlet boundary condition to the field on the positions
of the mirrors.
We focus our investigation on the discrete spectrum of created particles in a region of frequencies above $\omega_0$,
but considering an oscillation time $T=2L_0/c$, where $L_0$ is the initial and final length of the cavity.
Since for this oscillation duration a field mode, after perturbed by the moving mirror, never meets this mirror in motion again,
we exclude, in the creation of particles with frequencies above $\omega_0$, the effect of the re-interaction of a perturbed field mode with the mirror in a state of motion.
Then, we isolate the effect of the maximum speed of the mirror in creating particles with frequencies above $\omega_0$.
In this way, we describe, in a discrete spectrum of created particles,
the particle creation with frequencies above $\omega_0$ caused only by the relativistic aspect of the mirror's motion
(or the formation of relativistic bands).
In the literature, works have investigated the formation of these relativistic bands, but in the context of continuous spectra 
for single mirrors (see Lambrecht  \textit{et al.} \cite{Lambrecht-Jaekel-Reynaud-EPJD-1998}, Johansson \textit{et al.} \cite{ Johansson-PRL-2009} and Rego \textit{et al.} \cite{Rego-Alves-Alves-Farina-PRA-2013}).

The paper is organized as follows. 
In Sec. \ref{exact-formulas}, we present the model
to be investigated and write exact general formulas for the spectrum and total number of created particles in a dynamical cavity.
In Sec. \ref{application}, we make a brief check of the consistency of the exact formulas written in the previous section, 
comparing some of our results with analytical approximations found in the literature \cite{Dodonov-Klimov-PRA-1996}.
In Sec. \ref{spectrum}, we calculate the spectrum of created particles for 
an oscillation time $T=2L_0/c$, and discuss the appearance of relativistic bands.
In Sec. \ref{connect-cont-disc}, we investigate the connection and consistency between the relativistic band in the discrete spectrum 
(found in the previous section) and the relativistic band in a continuous spectrum for a relativistic oscillating single mirror 
found in the literature \cite{Lambrecht-Jaekel-Reynaud-EPJD-1998}.
In Sec. \ref{final}, we present a summary of our results and final comments.

\section{Exact formulas for the particle creation in a cavity}
\label{exact-formulas}
\begin{figure}[ht]
\begin{center}
\scalebox{0.25}{{\includegraphics{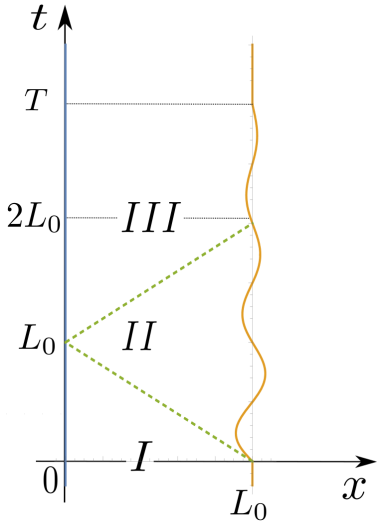}}}
\end{center}
\caption{\footnotesize{(color online). 
Trajectories of the mirrors (solid lines).
The static mirror is represented by the vertical solid (blue line) at $x=0$ .
The moving mirror, oscillating around $x=L_0$, with $L(t<0)=L(t>T)=L_0$, 
is represented by the orange solid line.
The dashed (green) lines are null lines separating region I from II, and region II from III. 
}}
\label{fig-law-of-motion-general}
\end{figure}

Let us start considering the massless scalar field in a two-dimensional spacetime satisfying the wave equation
(we assume throughout this paper $\hbar=c=1$)
\begin{equation}
\left(\partial _{t}^{2}-\partial _{x}^{2}\right) {\phi} \left(
t,x\right) =0,
\label{eq-Klein-gordon-massless}
\end{equation}
with the time-dependent boundary conditions
\begin{equation}
{\phi}\left( t,0\right)={\phi}\left[ t,L(t)\right]=0,
\label{boundary-condition}
\end{equation}
where $L(t)$ is an arbitrary prescribed law for the moving boundary with $L(t<0)=L(t>T)=L_0$,
where $L_0$ is the length of the cavity in the static situation, and $T$ is the time for which the boundary returns to its initial position $L_0$ (see Fig. \ref{fig-law-of-motion-general}).

Considering the procedure adopted by Moore \cite{Moore-1970}, Fulling and Davies \cite{Fulling-Davies-PRSA-1976}, 
the field in the cavity can be obtained by exploiting the conformal invariance of the wave equation (\ref{eq-Klein-gordon-massless}). 
The field solution, in the Heisenberg representation ${\phi}(t,x)$, is given by:
\begin{equation}
{\phi}(t,x)=\sum^{\infty }_{n=1}\left[
\hat{b}_{n}\phi_{n}\left( t,x\right) +H.c.\right],
\label{field-solution-I}
\end{equation} 
where the field modes $\phi_{n}(t,x)$ are given by:
\begin{eqnarray}
\phi_{n}(t,x)=\frac{i}{\sqrt{4n\pi}}\left[e^{-i n\pi R(v)} - e^{-in\pi R(u)}\right],
\label{field-solution-II}
\end{eqnarray} 
with $u=t-x$, $v=t+x$, and $R$ satisfying Moore's functional equation:
\begin{equation}
R[t+L(t)]-R[t-L(t)]=2.
\label{Moore-equation}
\end{equation}  
For $t<0$ (cavity in the static situation), $R(z)={z}/{L_{0}}$ and the field ${\phi}(t,x)$
can be written in terms of the complete set of function $\phi_{n}^{(0)}(t,x)$ as \cite{Dodonov-JMP-1993}:
\begin{equation}
{\phi}(t,x)=\sum^{\infty }_{n=1}\left[
\hat{b}_{n}\phi_{n}^{(0)}\left( t,x\right) +H.c.\right],
\label{field-solution-1}
\end{equation} 
where the field modes $\phi_{n}^{(0)}(t,x)$ are given by relation
\begin{eqnarray}
\phi_{n}^{(0)}(t,x)=\frac{i}{\sqrt{4n\pi}}\left[e^{-i n\pi v/L_0} - e^{-i n\pi u/L_0}\right],
\label{field-solution-2}
\end{eqnarray} 
with $[b_m,b_n^{\dagger}]=\delta_{mn}$.
Similarly to the equation (\ref{field-solution-1}), for $t>T$, when both boundaries are at rest again, 
the field solution ${\phi}(t,x)$ can be expanded as 
\begin{equation}
{\phi}(t,x)=\sum^{\infty }_{n=1}\left[
\hat{a}_{n}\phi_{n}^{(0)}\left( t,x\right) +H.c.\right].
\label{field-solution-3}
\end{equation} 
The new set of physical operators $(\hat{a},\hat{a}^{\dagger})$ is related to the old set $(\hat{b},\hat{b}^{\dagger})$ 
via the Bogoliubov transformation as
\begin{equation}
\hat{a}_{m}=\sum_{n=1}^{\infty}\{\hat{b}_{n}\alpha_{mn}+\hat{b}_{n}^{\dagger}\beta_{mn}^{*}\},
\end{equation} 
with the Bogoliubov coefficients given by \cite{Dodonov-Klimov-Manko-PLA-1990,Wegrzyn-MPLA-2004}
\begin{eqnarray}
\alpha_{mn}(t)&=&\frac{1}{2}\sqrt{\frac{m}{n}}\int_{t/L_{0}-1}^{t/L_{0}+1}dxe^{-i\pi\left[n R(L_{0}x) -  m x\right]},\nonumber\\
\beta_{mn}(t)&=&-\frac{1}{2}\sqrt{\frac{m}{n}}\int_{t/L_{0}-1}^{t/L_{0}+1}dxe^{-i\pi\left[n R(L_{0}x) + m x\right]},
\label{coef.bogoliubov}
\end{eqnarray}
where $R(z)$ is the solution of the Moore equation (\ref{Moore-equation}). The
unitarity condition for the Bogoliubov transformation is written as
$\sum_{n=1}^{\infty}\left[\left|\alpha_{mn}(t)\right|^2-\left|\beta_{mn}(t)\right|^2\right]=1$.
The number ${\cal{N}}_{n}(t)$ of created particles in the cavity, in a certain mode $n$
is given by:
\begin{equation}
{\cal{N}}_{n}(t)=\sum_{m=1}^{\infty}\left|\beta_{n m}(t)\right|^{2},
\label{N-n}
\end{equation}
with $\beta_{nm}(t)$ given by Eq. (\ref{coef.bogoliubov}). 
The total number ${\cal{N}}(t)$ of created particles in the cavity is given by:
\begin{equation}
{\cal{N}}(t)=\sum_{n=1}^{\infty}{\cal{N}}_{n}(t).
\label{N}
\end{equation}

Now, let us examine the cavity in the nonstatic situation ($t>0$). 
According to Cole and Schieve \cite{Cole-Schieve-PRA-1995}, the field modes in Eq. (\ref{field-solution-II}) are formed by left and
right-propagating parts. As causality requires, the field in
region I ($v\leq L_{0}$) (see Fig. \ref{fig-law-of-motion-general}) is not affected by the
boundary motion, so that, in this sense, this region is
considered as a ``static zone''. 
In region II ($v>L_{0}$ and $u\leq L_{0}$), the right-propagating parts of the field modes remain unaffected by the boundary motion, so that region II is also a static zone for these modes. On the other
hand, the left-propagating parts in region II are, in general,
affected by the boundary movement. 
In region III ($u>L_{0}$),
both the left and right-propagating parts are affected. In
summary, the functions corresponding to the left and right-propagating parts of the field modes are considered in the static zone if their argument $z$ ($z$ symbolizing $v$ or $u$) is such that $z\leq L_{0}$. 
For a certain spacetime point ($\tilde{t}$,$\tilde{x}$), the field operator ${\phi}(\tilde{t},\tilde{x})$ is known if its left and right-propagating parts, taken over, respectively, the null lines $v=z_{1}$ and $u=z_{2}$ (where $z_{1}=\tilde{t}+\tilde{x}$ and $z_{2}=\tilde{t}-\tilde{x}$), are known; or, in other words, ${\phi}(\tilde{t},\tilde{x})$ is known if $\left.R(v)\right|_{v=z_{1}}$ and $\left.R(u)\right|_{u=z_{2}}$ are known. 
Cole and Schieve \cite{Cole-Schieve-PRA-1995} proposed an 
elegant recursive method to obtain exactly the function $R$ for a general law of motion
of the boundary. The method consists in tracing back a sequence of null lines 
intersecting the worldline of the moving  
mirror at instants $t_i$, until, after a certain number $i=n$ of reflections, 
a null line traced back gets into the static zone, where the function $R$ is known. 
Following their procedure, one can write the solution of the Moore equation as
\cite{Cole-Schieve-PRA-1995,Alves-Granhen-Silva-Lima-PRD-2010,Alves-Granhen-CPC-2014}:
\begin{equation}
R(z)=2n(z)+[{z}-2\sum_{i=1}^{n(z)}L(t_i)]/L_0,
\label{R-cole}
\end{equation}
where $n$ is the number of reflections off the moving
boundary, necessary to connect the null line $t+x=z$ (or $t-x=z$) to a null line in the static zone. 
Using Eq. (\ref{R-cole}) in Eq. (\ref{N-n}) and (\ref{N}), 
we write the exact value for the number of created particles in the $r\text{th}$ mode (${\cal {N}}^{\left(r\right)}_{exa}$) and
the total number of created particles ${\cal{N}}_{exa}$, respectively, by
\begin{widetext}
\begin{equation}
{\cal {N}}^{\left(r\right)}_{exa}(t)=\sum_{s=1}^{\infty}\left|\frac{1}{2}\sqrt{\frac{r}{s}}\int_{t/L_{0}-1}^{t/L_{0}+1}dx  e^{-i\pi\left[s\left\{ 2n(L_{0}x)+[{L_{0}x}-2\sum_{i=1}^{n(L_{0}x)}L(t_{i})]/L_{0}\right\} +rx\right]}\right|^{2},
\label{N-n-exa}
\end{equation}
\end{widetext}
\begin{eqnarray}
{{\cal {N}}}_{exa}(t)=\sum_{r=1}^{\infty}{\cal {N}}^{\left(r\right)}_{exa}(t).
\label{N-exa}
\end{eqnarray}

The formulas (\ref{N-n-exa}) and (\ref{N-exa}) are valid for
an arbitrary prescribed law $L(t)$ for the moving boundary, provided that $L(t<0)=L(t>T)=L_0$.
In the next sections we will apply these formulas to the following class of laws of motion for the moving mirror: 
\begin{equation}
 L(t) = 
  \begin{cases} 
   L_0,\; t<0 \\
	 L_{0}+a\sin\left(2\pi t/l_{0}\right),\; 0\leq t \leq T, \\
   L_0,\;t>T
  \end{cases}
	\label{law-of-motion}
\end{equation}
where $a>0$ is the amplitude of oscillation, and $l_0$ needs to be chosen appropriately
so that $L(t)=L_0$ for $t>T$. Along the text, we consider $\omega_0=2\pi/l_{0}$ as the frequency of oscillation of the moving mirror.

\section{Comparison with approximate analytical results}
\label{application}

As a first application of our computations based on the exact formulas (\ref{N-n-exa}) and (\ref{N-exa}), we compare
some of our results for the total number of created particles with those obtained by analytical approximations found in the literature \cite{Dodonov-Klimov-PRA-1996}.
Let us consider a particular resonant law of motion typically considered in the investigation
of the DCE \cite{Dodonov-JMP-1993,Dodonov-Klimov-PRA-1996}, given by Eq. (\ref{law-of-motion}) with
$a=\varepsilon L_0$, $l_0=L_0$, being $\varepsilon>0$, and $\varepsilon L_{0}$ the amplitude of oscillation.
Note that in this particular case the frequency $\omega_0=2\pi/L_{0}$  is twice the frequency of the first quantum mode, $\pi/L_{0}$, inside the static cavity of length $L_0$. This law of motion leads to a resonant particle creation in the cavity.
Dodonov and Klimov \cite{Dodonov-Klimov-PRA-1996}, considering this law of motion in the context of non-relativistic velocities and low amplitudes, obtained perturbatively the approximate average total number of particles created, ${\cal{N}}_{app}$, as given by
\begin{eqnarray}
{\cal{N}}_{app}(T)=\frac{1}{\pi^2}\left[\left(1-\frac{\kappa^2}{2}\right)K^2(\kappa)-E(\kappa)K(\kappa)\right],  
\label{N-app}
\end{eqnarray}
where $K(\kappa)$ and $E(\kappa)$ are the complete elliptic integrals of the first and second kind, respectively, and
$\kappa=\sqrt{1-e^{-4{\varepsilon\pi T}/{L_{0}}}}$.
The authors also obtained the following formula for the number ${\cal{N}}_{app}^{(1)}$ of created particles in the first
(fundamental) mode of the cavity:
\begin{eqnarray}
{\cal{N}}_{app}^{(1)}(T)=\frac{2}{\pi^2}K(\kappa)E(\kappa)-\frac{1}{2}.  
\label{N-1-app}
\end{eqnarray}
The results in Eq. (\ref{N-app}) and (\ref{N-1-app}) were considered valid in the limit $\varepsilon\ll 1$ \cite{Dodonov-Klimov-PRA-1996}.

\begin{figure}[ht]
\begin{center}
\scalebox{0.6}{{\includegraphics{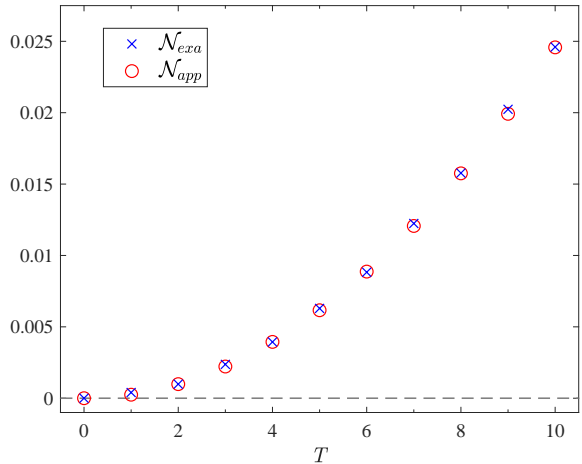}}}
\end{center}
\caption{\footnotesize{(color online). Comparison between the number of particles (vertical axis)
versus $T$ (horizontal axis), via approximate formula ${\cal{N}}_{app}$ (circles) and exact formula ${\cal{N}}_{exa}$ (crosses),
for the law of motion given in Eq. (\ref{law-of-motion}), with $\varepsilon=10^{-2}$ and  $L_0=1$ ($\omega_0=2\pi$).
The dashed line serves as a reference for the zero value of the number of particles.
}}
\label{fig-N-L0-1-epsilon-0-01}
\end{figure}

Let us compare the results for the total number of particles, using the formulas ${\cal{N}}_{exa}$ [Eq. (\ref{N-exa})] 
and ${\cal{N}}_{app}$ [Eq. (\ref{N-app})].
We start this comparison examining the case with $\varepsilon=10^{-2}$, which implies
in a maximum velocity $v$ of the mirror such that $v\approx 0.06$. 
In Fig. \ref{fig-N-L0-1-epsilon-0-01}, 
corresponding to the case with $L_0=1$ ($\omega_0=2\pi$),
one can see an agreement between ${\cal{N}}_{app}$ (circles) and ${\cal{N}}_{exa}$ (crosses).
In addition, both results are in agreement with numerical ones found by Ruser \cite{Ruser-JOptB-2005}
(other numerical approaches to solve DCE problems have also been developed 
\cite{Antunes-hepph-2003,Ruser-JPA-2006, Lombardo-Mazzitelli-Soba-Villar-PRA-2016,Villar-PRA-2017,Villar-Soba-PRE-2017}). 
We also verified agreement between ${\cal{N}}_{exa}$ [Eq. (\ref{N-exa})] and ${\cal{N}}_{app}$ [Eq. (\ref{N-app})]
for $s<-2$ in $\varepsilon=10^{s}$.

\begin{figure}[ht]
\begin{center}
\scalebox{0.6}{{\includegraphics{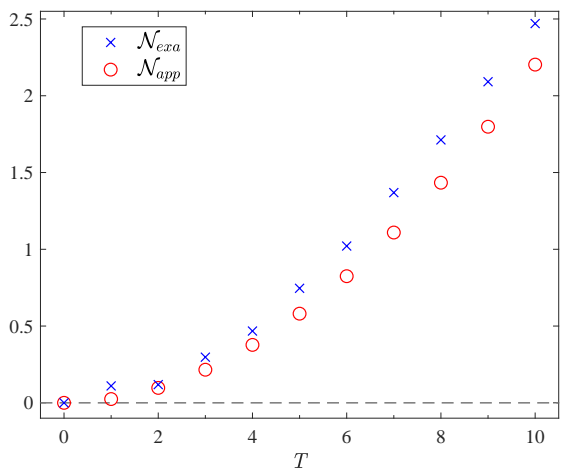}}}
\end{center}
\caption{\footnotesize{(color online). Comparison between the number of particles (vertical axis)
versus $T$ (horizontal axis), via approximate formula ${\cal{N}}_{app}$ (circles) and exact formula ${\cal{N}}_{exa}$ (crosses),
for the law of motion given in Eq. (\ref{law-of-motion}), with $\varepsilon=10^{-1}$ and  $L_0=1$ ($\omega_0=2\pi$).
The dashed line serves as a reference for the zero value of the number of particles.
}}\label{fig-N-L0-1-epsilon-0-1}
\end{figure}

Now, let us investigate the case with $\varepsilon=10^{-1}$, which means 
a maximum velocity $v\approx 0.6$. 
In Fig. \ref{fig-N-L0-1-epsilon-0-1}, one can see a certain disagreement
between ${\cal{N}}_{exa}$ [Eq. (\ref{N-exa})] and  
${\cal{N}}_{app}$ [Eq. (\ref{N-app})], with
${\cal{N}}_{exa}>{\cal{N}}_{app}$ and
${\cal{N}}_{exa}-{\cal{N}}_{app}$ growing in time.
It is worth mentioning that a similar disagreement (for $\epsilon=10^{-1}$) 
was also observed in the literature \cite{Ruser-JOptB-2005},
when values calculated via numerical methods were compared 
to values from ${\cal{N}}_{app}$ [Eq. (\ref{N-app})].
This indicates that the disagreement found in Fig. \ref{fig-N-L0-1-epsilon-0-1}
reveals not a failure in predictions based on ${\cal{N}}_{exa}$ [Eq. (\ref{N-exa})],
but a limit of validity for ${\cal{N}}_{app}$ [Eq. (\ref{N-app})]
(namely, ${\cal{N}}_{app}$ is better valid for $s\leq-2$ in $\varepsilon=10^{s}$). 

\section{Spectrum of created particles}
\label{spectrum}

In a cavity with length $L_0$, with one of the mirrors in motion (for instance, the right one), the field modes, after perturbed by the right oscillating mirror, are reflected by the left (static) mirror and go back to the right mirror again.
If the field modes return to the right mirror and find it still in motion, the perturbed field modes undergo a new perturbation
(re-interaction). 
On the other hand, if the perturbed field modes find the right mirror at rest, they are simply reflected, going in the opposite direction but with no new perturbation added to them.
When re-interactions are allowed in an oscillating cavity (what happens when $T>2L_0$),
even with non-relativistic velocities, particles can be produced with frequencies higher than the oscillation frequency \cite{Lambrecht-Jaekel-Reynaud-EPJD-1998,Dodonov-Klimov-PRA-1996,
Alves-Farina-Granhen-PRA-2006}.
For instance, for the law of motion in Eq. (\ref{law-of-motion}), 
with $l_0=L_0$, $a=10^{-8} L_0$ (which means $v=2\pi\times 10^{-8}$) and $T>2L_0$, particles can be created with frequencies $(2n+1)\pi/L_0$ ($n=0,1,2...$) and, for $n>0$, with frequencies higher than $\omega_0=2\pi/L_0$ \cite{Dodonov-Klimov-PRA-1996,Alves-Farina-Granhen-PRA-2006}.
Then, the results in the literature \cite{Dodonov-Klimov-PRA-1996,Alves-Farina-Granhen-PRA-2006} show that the particle creation via DCE in the resonant cavity described by (\ref{law-of-motion}), with $T>2L_0$, can be characterized by: a discrete spectrum; the possibility of several re-interactions of the field modes with the moving mirror; and particles produced with frequencies higher than the oscillation frequency even with a non-relativistic moving mirror.

\begin{figure}[ht]
\begin{center}
\scalebox{0.3}{{\includegraphics{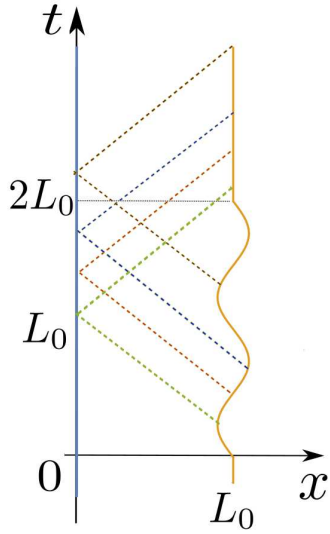}}}
\end{center}
\caption{\footnotesize{(color online). 
Trajectories of the mirrors (solid lines).
The static mirror is represented by the vertical solid (blue line) at $x=0$ .
The trajectory of the moving mirror, according to Eq. (\ref{law-of-motion})
(with $l_0=L_0$ and $T=2L_0$), is given by the orange line.
The dashed lines represent some null lines related to the field modes perturbed
by the moving mirror. 
Note that, after reflected by the left static mirror, all field modes,
which were perturbed by the right mirror in motion, find the right mirror again already at rest.
}}
\label{fig-law-of-motion-2-L0}
\end{figure}

In the present section we will focus on oscillatory motions obeying
Eq. (\ref{law-of-motion}), with $l_0=L_0$ and $T=2L_0$ (see Fig. \ref{fig-law-of-motion-2-L0}). 
This enables us to exclude the effect of the re-interaction of a perturbed field mode with the mirror in a state of motion, 
so that we can isolate only the role of the maximum speed of the mirror in creating particles with frequency above $\omega_0$
(relativistic band).
In this way, the relativistic band can be assigned exclusively to the relativistic aspect of the velocity of the mirror.
(as occurs for a relativistic single mirror \cite{Lambrecht-Jaekel-Reynaud-EPJD-1998}).
This motion law is interesting because all field modes perturbed by the moving 
(right) mirror, after reflected on the static (left) mirror at $x=0$, go
back to the right, but find the right mirror at rest. 
This is illustrated by the dashed lines in Fig. \ref{fig-law-of-motion-2-L0}, 
which represent null lines related to the field modes
perturbed by the right mirror in motion. 
In this manner, for the law of motion in Eq. (\ref{law-of-motion}) with $T=2L_0$, the
values of ${\cal {N}}^{\left(r\right)}_{exa}$ and ${\cal{N}}_{exa}$ 
exclude the effect of a new interaction of the perturbed field modes with the right mirror in the state of motion.

\begin{figure}[ht]
\begin{center}
\scalebox{0.6}{{\includegraphics{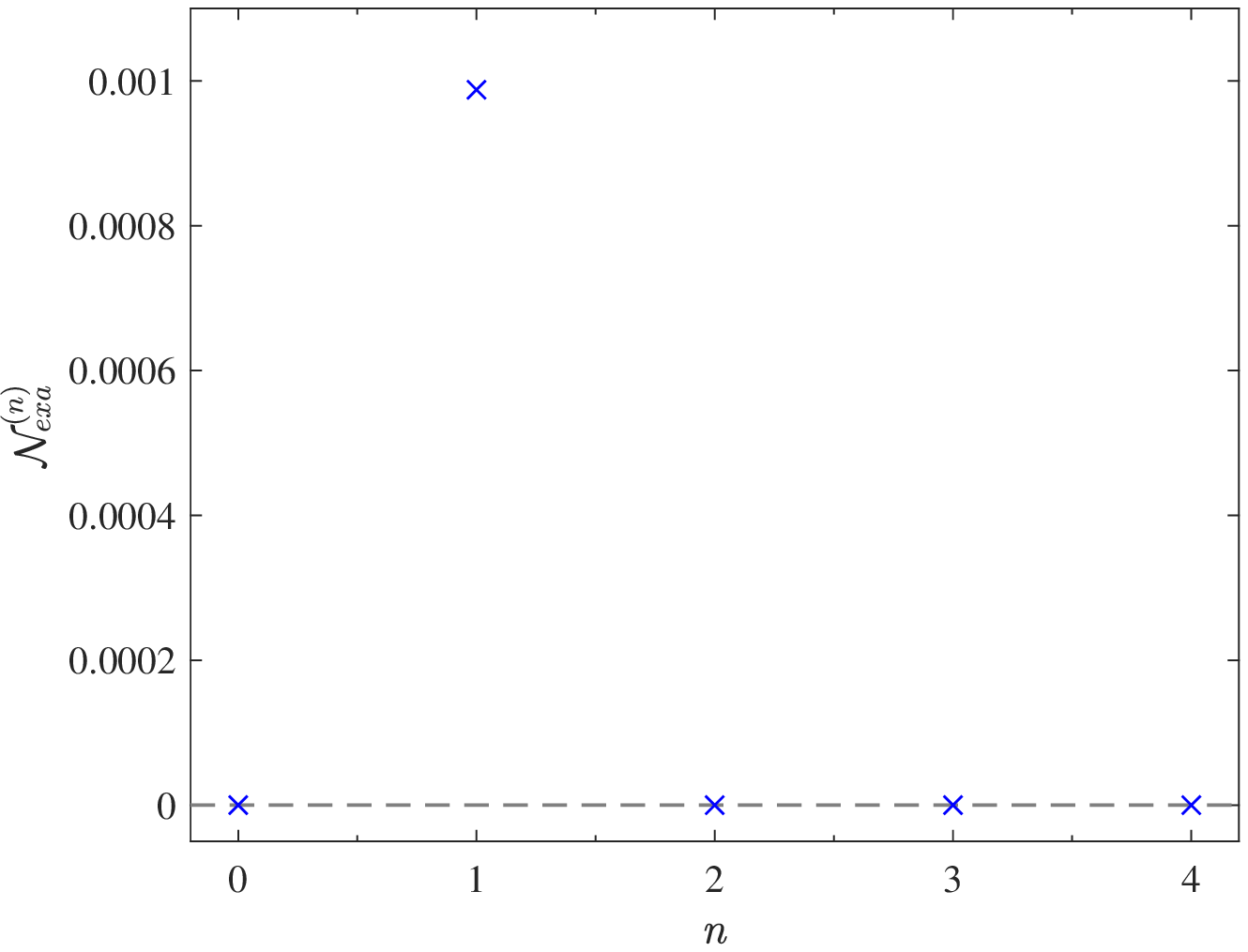}}}
\end{center}
\caption{\footnotesize{The number of created particles ${\cal {N}}^{\left(n\right)}_{exa}$ (vertical axis) versus
$n=\omega L_0/\pi$ (horizontal axis), for the law of motion given in Eq. (\ref{law-of-motion}), with 
$T=2L_0$, $a=\varepsilon L_0$, $L_0=l_0=1$ and $\varepsilon=10^{-2}$ 
(maximum velocity $v\approx 0.06$).
The dashed line serves as a reference for the value ${\cal {N}}^{\left(n\right)}_{exa}=0$.
Note that $n=1$ represents half of the oscillation frequency,
and $n=2$ indicates the oscillation frequency $\omega_0$.
One can see no creation of particles with frequencies larger than $\omega_0=2\pi$ ($n=2$).
}}
\label{fig-N-n-L0-1-epsilon-0-01}
\end{figure}

Considering the law of motion in Eq. (\ref{law-of-motion}), with 
$T=2L_0$, $a=\varepsilon L_0$, $l_0=L_0$, $\varepsilon=10^{-2}$ ($v\approx 0.06$), 
and using the exact formula (\ref{N-n}), we obtain that there is 
no effective creation of particles with frequency above the oscillation frequency of the cavity ($2\pi/L_0$),
with the particle creation restricted to
the fundamental mode $n=1$ ($\pi/L_0$) (see Fig. \ref{fig-N-n-L0-1-epsilon-0-01}),
which has half of the oscillating frequency $\omega_0$. 
This is in agreement with the approximate results 
found in the literature \cite{Dodonov-Klimov-PRA-1996}.
The expected number of particles obtained
by us [via Eq. (\ref{N-n-exa})] for the first mode is in agreement with that 
obtained via approximate [Eq. (\ref{N-1-app})]:
${\cal{N}}_{exa}^{(1)}\approx {\cal{N}}_{app}^{(1)} \approx 0.001$. 

\begin{figure}[ht]
\begin{center}
\scalebox{0.6}{{\includegraphics{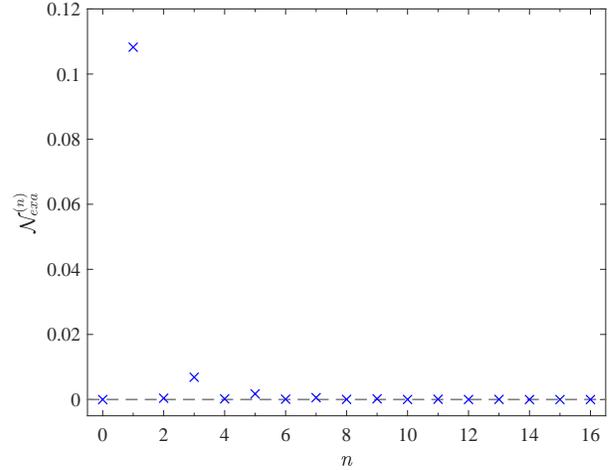}}}
\end{center}
\caption{\footnotesize{The number of created particles ${\cal {N}}^{\left(n\right)}_{exa}$ (vertical axis) versus
$n=\omega L_0/\pi$ (horizontal axis), for the law of motion given in Eq. (\ref{law-of-motion}), with 
$T=2L_0$, $a=\varepsilon L_0$, $L_0=l_0=1$ and $\varepsilon=10^{-1}$
(maximum velocity $v\approx 0.6$).
The dashed line serves as a reference for the value ${\cal {N}}^{\left(n\right)}_{exa}=0$.
Note that $n=1$ represents half of the oscillation frequency,
and $n=2$ indicates the oscillation frequency $\omega_0$.
For $n=3$ and $n=5$, one can see the creation of particles with frequencies larger than
$\omega_0=2\pi$ ($n=2$).
}}
\label{fig-N-n-L0-1-epsilon-0-1}
\end{figure}

Now, considering $\varepsilon=10^{-1}$ ($v\approx 0.6$) in Eq. (\ref{law-of-motion}) (with $T=2L_0$ and $l_0=L_0$),
the exact method used here [Eq. (\ref{N-n-exa})] also predicts, beyond creation in the fundamental mode  $n=1$ ($\pi/L_0$), 
particle creation in an additional band (frequencies larger than $\omega_0=2\pi/L_0$).
For instance, in Fig. \ref{fig-N-n-L0-1-epsilon-0-1} one can see the creation of particles with frequencies $3\pi/L_0$ (mode $n=3$) 
and $5\pi/L_0$ ($n=5$). 
Since this particle creation with frequencies above $\omega_0$ is not related to the re-interactions of perturbed field modes with the right mirror in a state of motion, but caused only by the relativistic aspect (in this case, $v\approx 0.6$) of the mirror's motion,
this region of frequencies with ($\omega>\omega_0$) is called a relativistic band \cite{Rego-Alves-Alves-Farina-PRA-2013}.

\begin{figure}[ht]
\begin{center}
\scalebox{0.6}{{\includegraphics{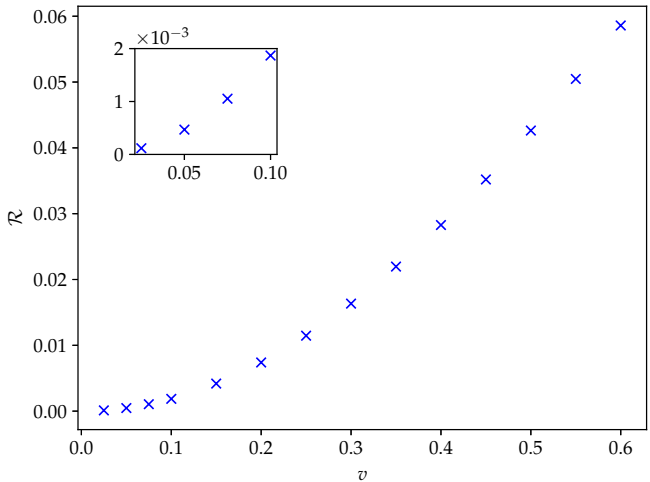}}}
\end{center}
\caption{\footnotesize{The ratio ${\mathcal{R}}={\cal {N}}^{\left(3\right)}_{exa}/{\cal {N}}^{\left(1\right)}_{exa}$
(vertical axis) versus $v=2\pi a$ (horizontal axis), where $a$
is the amplitude of oscillation given in the law of motion in Eq. (\ref{law-of-motion}), with 
$L_0=l_0=1$.
}}
\label{fig-ratio-N3-N1-evolution}
\end{figure}

To estimate the relevance of the relativistic band as the maximum velocity of oscillation increases,
we consider Eq. (\ref{law-of-motion})
with $L_0=l_0=1$, and $a$ ($v=2\pi a$) varying
from $0$ ($v=0$) to $0.1$ ($v\approx 0.6$).
In Fig. \ref{fig-ratio-N3-N1-evolution}, we show [using Eq. (\ref{N-n-exa})] the behaviour of the 
ratio ${\mathcal{R}}={\cal {N}}^{\left(3\right)}_{exa}/{\cal {N}}^{\left(1\right)}_{exa}$ 
as a function of $v$. 
We highlight the following results: 
$v \approx 2 \pi \times 10^{-3}$ $\Rightarrow$ ${\mathcal{R}}\approx 7.4\times 10^{-6}$;
$v \approx 2 \pi \times 10^{-2}$ $\Rightarrow$ ${\mathcal{R}}\approx 7.4 \times 10^{-4}$
(these values correspond to the case shown in Fig. \ref{fig-N-n-L0-1-epsilon-0-01},
and the low value of ${\mathcal{R}}$ explains the null visualization of a relativistic band);
$v \approx 10^{-1}$ $\Rightarrow$ ${\mathcal{R}}\approx 1.9 \times 10^{-3}$
(this velocity is the maximum effective velocity considered  by Wilson \textit{et al.} 
in the first observation of the DCE \cite{Wilson-Nature-2011});
$v \approx 3.0\times 10^{-1}$  $\Rightarrow$ ${\mathcal{R}}\approx 1.6 \times 10^{-2}$
(this velocity is the maximum effective velocity considered in the experiment by Schneider \textit{et al.}\cite{Schneider-et-al-PRL-2020});
$v \approx 2 \pi \times 10^{-1}$ $\Rightarrow$ ${\mathcal{R}}\approx 6.3 \times 10^{-2}$
(these values correspond to the relativistic band visualized in Fig. \ref{fig-N-n-L0-1-epsilon-0-1}).

The increase of ${\mathcal{R}}$ with $v$, shown in Fig. \ref{fig-ratio-N3-N1-evolution}, 
describes the formation of relativistic bands or, in other words, a significant particle creation with frequencies above $\omega_0$
caused only by the relativistic aspect of the mirror's motion.

\section{Connecting discrete and continuous relativistic bands}
\label{connect-cont-disc}

In the present section, we investigate the connection between the relativistic band in the discrete spectrum shown in Fig. \ref{fig-N-n-L0-1-epsilon-0-1} and the relativistic band in a continuous spectrum for a relativistic oscillating single mirror \cite{Lambrecht-Jaekel-Reynaud-EPJD-1998}.

Let us start the investigation examining the spectrum shown in Fig. \ref{fig-N-n-L0-1-epsilon-0-01},
where one can see that the creation of particles occurs only for the frequency $\omega_0/2$ and, consequently, there is no creation of particles with frequency beyond $\omega_0$.
Although Fig. \ref{fig-N-n-L0-1-epsilon-0-01} does not look like a parabola, the result shown in this figure is deeply connected to the parabolic continuous spectrum of a single moving mirror with non-relativistic velocities \cite{Lambrecht-Jaekel-Reynaud-PRL-1996}, for which
the spectral distribution has a maximum at $\omega_0/2$ and there is no particle creation with frequencies higher than $\omega_0$.
To clarify this connection, let us compare the cases of cavities with the moving mirror oscillating 
according to Eq. (\ref{law-of-motion}), with a fixed frequency $\omega_0=2\pi$
[in other words a fixed value $l_0=1$], fixed amplitude of oscillation $a$, 
but with different values of $L_0$, with $T=2L_0$.
We reinforce that, since we are considering the condition $T=2L_0$, all field modes, after perturbed by the oscillating right mirror and reflected by the left static mirror, do not find  again the right mirror in a state of motion.
We also remark that this is an important condition in order to make the transition 
from a discrete spectrum to a continuous one (produced by a single moving mirror and discussed in the literature \cite{Lambrecht-Jaekel-Reynaud-PRL-1996}), since the field modes, after perturbed by an oscillating single mirror, go to infinity and never interact with the moving mirror again.

\begin{figure}[ht]
\begin{center}
\scalebox{0.6}{{\includegraphics{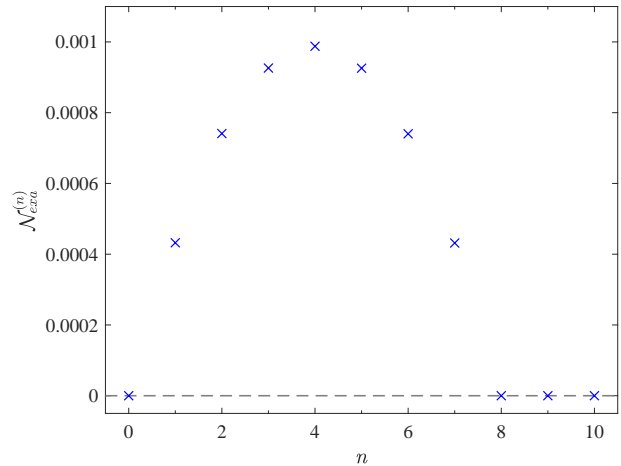}}}
\end{center}
\caption{\footnotesize{The number of created particles ${\cal {N}}^{\left(n\right)}_{exa}$ (vertical axis) versus
$n=\omega L_0/\pi$ (horizontal axis), for the law of motion given in Eq. (\ref{law-of-motion}), with $L_0=4$, $l_0=1$ and $a=10^{-2}$
(maximum velocity $v=2\pi a\approx 0.06$).
The dashed line serves as a reference for the value ${\cal {N}}^{\left(n\right)}_{exa}=0$.
Note that $n=4$ represents half of the oscillation frequency,
and $n=8$ indicates the oscillation frequency $2\pi$.
One can see that there is no creation of particles with frequency larger than $2\pi$ ($n=8$).
}}
\label{fig-limit-0-01}
\end{figure}

For the law of motion in Eq. (\ref{law-of-motion}),
with $a=10^{-2}$, $l_0=1$ ($\omega_0=2\pi$ and $v\approx 0.06$)
and $T=2L_0$, we have for $L_0=1$ and $L_0=4$ the results shown
in Fig. \ref{fig-N-n-L0-1-epsilon-0-01} and Fig. \ref{fig-limit-0-01}, respectively.
One can see that, with the increase in $L_0$ [for instance from $L_0=1$ (Fig. \ref{fig-N-n-L0-1-epsilon-0-01}) 
to $L_0=4$ (Fig. \ref{fig-limit-0-01})], maintaining the same oscillation frequency $2\pi$, it occurs a population of particle creation in several frequencies (smaller than $\omega_0$) in addition to $\omega_0/2=\pi$.
The discrete values obtained outline a parabolic spectrum (Fig. \ref{fig-limit-0-01}), with the maximum number of created particles with frequency $\pi$ and no particles created with frequency higher than $\omega_0=2\pi$.
This is in accordance with the predictions found in the literature for a continuous spectrum for a single oscillating mirror 
\cite{Lambrecht-Jaekel-Reynaud-PRL-1996}.
In other words, the spectrum shown in Fig. \ref{fig-N-n-L0-1-epsilon-0-01} 
is a germinal version of a spectrum with a parabolic shape,
in the sense that, as $L_0$ is increased (but keeping the same frequency value), 
more and more the discrete spectrum outlines a continuous parabolic one.
This reveals a consistency between the results obtained here 
[for the discrete spectra in a cavity provided by the exact  formula (\ref{N-n-exa})]
and those found in the literature \cite{Lambrecht-Jaekel-Reynaud-PRL-1996},
for a continuous spectra for a non-relativistic oscillating single mirror. 

\begin{figure}[ht]
\begin{center}
\scalebox{0.6}{{\includegraphics{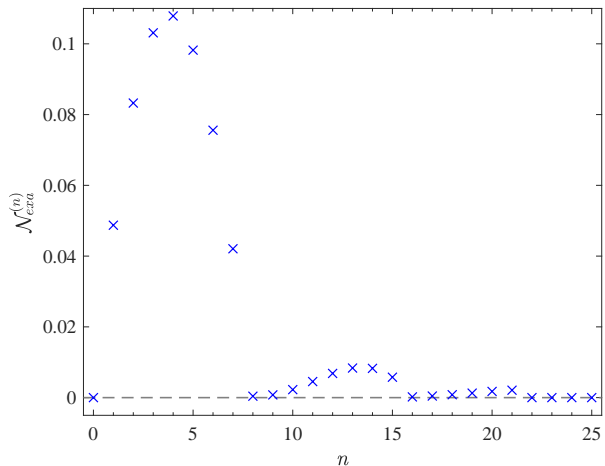}}}
\end{center}
\caption{\footnotesize{The number of created particles ${\cal {N}}^{\left(n\right)}_{exa}$ (vertical axis) versus
$n=\omega L_0/\pi$ (horizontal axis), for the law of motion given in  Eq. (\ref{law-of-motion}), with $L_0=4$, $l_0=1$ and $a=10^{-1}$
(maximum velocity $v=2\pi a \approx 0.6$).
The dashed line serves as a reference for the value ${\cal {N}}^{\left(n\right)}_{exa}=0$.
Note that $n=4$ represents half of the oscillation frequency,
and $n=8$ the oscillation frequency $2\pi$.
One can see the creation of particles with frequencies larger than
$\omega_0=2\pi$ ($n=8$).
}}
\label{fig-limit-0-1}
\end{figure}

Now, we continue our investigation examining the spectrum shown
in Fig. \ref{fig-N-n-L0-1-epsilon-0-1} (maximum velocity $v\approx 0.6$), where one can see the creation of
particles with frequency $\omega_0/2$, and also creation of particles with frequencies above the oscillating frequency $\omega_0$,
for instance $3\omega_0/2$ and $5\omega_0/2$, but vanishing for all frequencies $\omega$ equal to an integer multiple of $\omega_0$.
Although one can say that Fig. \ref{fig-N-n-L0-1-epsilon-0-1} does not look like a succession of arches, the result shown in that figure is connected to the continuous spectrum for a relativistic oscillating single mirror, formed by a succession of arches, 
each one limited by two successive multiples of $\omega_0$, and vanishing for all frequencies $\omega$ equal to an integer multiple of $\omega_0$ \cite{Lambrecht-Jaekel-Reynaud-EPJD-1998}.
To clarify this connection, let us consider the law motion in Eq. (\ref{law-of-motion}),
with $a=10^{-1}$, $l_0=1$ ($\omega_0=2\pi$ and $v\approx 0.6$),
and $T=2L_0$. 
We show in Fig. \ref{fig-N-n-L0-1-epsilon-0-1} and Fig. \ref{fig-limit-0-1} 
the results for $L_0=1$ and $L_0=4$, respectively.
Increasing $L_0$, for instance from $L_0=1$ (Fig. \ref{fig-N-n-L0-1-epsilon-0-1}) to $L_0=4$ (Fig. \ref{fig-limit-0-1}), it occurs a population of particles created in several other frequency modes in addition to $\pi$, outlining a continuous spectrum formed by a succession of arches (Fig. \ref{fig-limit-0-1}), each one limited by two successive multiples of $\omega_0$, and vanishing for all frequencies $\omega$ equal to an integer multiple of $\omega_0$, in connection with the predictions found in the literature for a continuous spectrum
for a single relativistic oscillating mirror \cite{Lambrecht-Jaekel-Reynaud-EPJD-1998}.
The spectrum shown in Fig. \ref{fig-N-n-L0-1-epsilon-0-1} 
is then an initial version of a spectrum with a succession of arches, in the sense that, 
as $L_0$ is increased, more and more the discrete spectrum outlines a continuous succession of arches, 
exhibiting additional (relativistic) bands with frequencies higher then $\omega_0$ \cite{Lambrecht-Jaekel-Reynaud-EPJD-1998}.
Again, this reveals a consistency between the results for discrete spectra (cavity) provided by the exact formula (\ref{N-n-exa}) and those for continuous spectra for a relativistic oscillating single mirror found in the literature \cite{Lambrecht-Jaekel-Reynaud-EPJD-1998}.

\section{Final remarks}
\label{final}
In the present paper, we investigated the formation,
via dynamical Casimir effect, of relativistic bands in the discrete spectrum of created particles in an oscillating one-dimensional 
resonant cavity.
We considered a real scalar field obeying
Eq. (\ref{eq-Klein-gordon-massless}), under the boundary conditions given in Eq. (\ref{boundary-condition}). 
We wrote, based on previous works in the literature \cite{Dodonov-Klimov-Manko-PLA-1990,Wegrzyn-MPLA-2004,Cole-Schieve-PRA-1995}, 
exact formulas for the spectrum [Eq. (\ref{N-n-exa})] and total number of created particles [Eq. (\ref{N-exa})].
Although these formulas are valid for an arbitrary prescribed law of motion for the oscillating mirror,
we put our attention on the class of laws of motion given by Eq. (\ref{law-of-motion}),
and, more specifically, considering $a=\varepsilon L_0$, $l_0=L_0$ and $T=2L_0/c$. 

With the first two choices ($a=\varepsilon L_0$ and $l_0=L_0$), Eq. (\ref{law-of-motion}) 
describes a resonant law of motion typically investigated
in the context of the DCE \cite{Dodonov-JMP-1993,Dodonov-Klimov-PRA-1996},
where the oscillation frequency is $\omega_0=2\pi/L_{0}$ 
(twice the frequency of the first mode $\pi/L_{0}$).
In addition, the choice of the time
of oscillation $T=2L_0/c$ is such that a field mode, after perturbed by the moving mirror, never meets this mirror in motion again
(see Fig. \ref{fig-law-of-motion-2-L0}).
This allowed us to exclude the effect of the re-interaction of a perturbed field mode with the mirror in a state of motion, 
so that we could isolate only the role of the maximum speed of the mirror in creating particles with frequency above $\omega_0$.

Using Eq. (\ref{N-n-exa}), we computed the spectrum of created particles
when $v\approx 0.06 c$ (Fig. \ref{fig-N-n-L0-1-epsilon-0-01}) and 
$v\approx 0.6 c$ (Fig. \ref{fig-N-n-L0-1-epsilon-0-1}).
In Fig. \ref{fig-N-n-L0-1-epsilon-0-01}, we got 
no visible creation of particles with frequency above $\omega_0=2\pi/L_0$ 
(or no visualization of a relativistic band),
with creation of particles restricted to the fist mode $n=1$.
More precisely, the relativistic band exists, but the number of particles is, for the 
mode $n=3$, only approximately $7.4 \times 10^{-4}$ of the number of created particles in the first mode.
In Fig. \ref{fig-N-n-L0-1-epsilon-0-1}, we can
visualize an effective creation of particles with frequency above $\omega_0=2\pi/L_0$.
In this case, the relativistic band is such that the number of particles for the mode $n=3$ 
is approximately $6.3 \times 10^{-2}$ of the number of created in the first mode.
In Fig. \ref{fig-ratio-N3-N1-evolution}, corresponding to Eq. (\ref{law-of-motion}) with $L_0=l_0=1$, 
we describe the enhancement of the relativistic band in a discrete spectrum of created particles 
as the maximum velocity of oscillation increases.
Finally, we showed the connection between the relativistic band in the discrete spectrum shown in Fig. \ref{fig-N-n-L0-1-epsilon-0-1} 
and a relativistic band in a continuous spectrum (outlined in Fig. \ref{fig-limit-0-1}) for a relativistic oscillating single mirror \cite{Lambrecht-Jaekel-Reynaud-EPJD-1998}. 

\begin{acknowledgments}

The authors thank Alessandra N. Braga, Amanda E. da Silva, Andreson L. C. Rego, 
Edson C. M. Nogueira, Jeferson D. Lima Silva and Van S\'{e}rgio Alves
for careful reading of this paper, fruitful discussions and suggestions.
D.T.A. thanks the hospitality of the Centro de F\'{i}sica, Universidade do Minho, Braga, Portugal.
E.R.G. thanks the hospitality of the Programa de P\'{o}s-Gradua\c{c}\~{a}o em F\'{i}sica, 
Universidade Federal do Par\'{a}, Bel\'{e}m, Par\'{a}, Brazil.
\end{acknowledgments}
%

%

\end{document}